\documentclass[twocolumn,showpacs,preprintnumbers,amsmath,amssymb]{revtex4}
\usepackage{epsf,epsfig,graphics,graphicx,colordvi}

\usepackage{dcolumn}
\usepackage{bm}

\begin{document}

\preprint{\ }

\title{Low-Redshift Cosmic Baryon Fluid on Large Scales and
She-Leveque Universal Scaling}

\author{Ping He$^1$, Jiren Liu$^2$, Long-Long Feng$^3$, Chi-Wang Shu$^4$,
     Li-Zhi Fang$^5$}

\affiliation{$^1$ Institute of Theoretical Physics, Chinese
Academy of Sciences, Beijing 100080, China \\
$^2$ Center for Astrophysics, University of Science and Technology
  of China, Hefei 230026, China\\
$^3$Purple Mountain Observatory, Nanjing 210008, China \\
$^4$ Division of Applied Mathematics, Brown University,
Providence, RI 02912 \\
$^5$ Department of Physics, University of Arizona, Tucson, AZ
85721}
\date{\today}

\begin{abstract}

We investigate the statistical properties of cosmic baryon fluid
in the nonlinear regime, which is crucial for understanding the
large-scale structure formation of the universe. With the
hydrodynamic simulation sample of the Universe in the cold dark
matter model with a cosmological constant, we show that the
intermittency of the velocity field of cosmic baryon fluid at
redshift $z=0$ in the scale range from the Jeans length to about
16 h$^{-1}$ Mpc can be extremely well described by She-Leveque's
universal scaling formula. The baryon fluid also possesses the
following features: (1) for volume weight statistics, the
dissipative structures are dominated by sheets, and (2) the
relation between the intensities of fluctuations is hierarchical.
These results imply that the evolution of highly evolved cosmic
baryon fluid is similar to a fully developed turbulence.

\end{abstract}


\maketitle

{\it Introduction.}---The formation and evolution of large-scale
structure of the universe is governed by the gravitational
clustering of cosmic matter, in which about 72\% is dark energy,
24\% cold dark matter, and 4\% baryon. Most baryonic matter is in
the form of gas. Therefore, the evolution of the density and
velocity fluctuations of cosmic baryon fluid is dominated by the
underlying gravitational potential of dark matter. In the linear
regime, the baryon fluid follows the mass and velocity fields of
collisionless dark matter point by point. In the nonlinear regime,
however, as first pointed out by Shandarin and Zeldovich in their
early study of structure formation, the dynamical behavior of
cosmic matter clustering on large scales is similar to turbulence
\cite{SZ}. They emphasized that the motion of self-gravitation
matter in the expanding universe is like that of noninteracting
matter moving by inertia. In other words, cosmic matter underwent
a scale-free evolution somewhat like fully developed turbulence in
inertial range.

Nevertheless, the dynamical difference between the structure
formation of cosmic matter and turbulent flow in incompressible
fluid is obvious. The latter is rotational in general \cite{LL},
while the former is irrotational, because vorticities do not grow
during the clustering \cite{Pe}. In turbulence, energy passes from
large to the smallest eddies, and finally dissipates into thermal
motion, while cosmic baryonic gas falls into massive halos to form
structures, including light-emitting objects.

Yet, the turbulencelike behavior of cosmic baryon fluid has been
gradually noticed in the last decade. Although the evolution of
cosmic baryon fluid is governed by the Naiver-Stokes equation, the
dynamics of growth modes of the fluid can be sketched by a
stochastic force driven Burgers' turbulence \cite{Jo}
\begin{equation}
\frac{\partial \varphi}{\partial t}- \frac{1}{2a^2}(\nabla \varphi)^2 -
\frac{\nu}{a^2}\nabla^2 \varphi =\phi,
\end{equation}
where $\varphi$ is velocity potential, ${\bm
v}=(1/a)\nabla\varphi$, $a$ is the scale function of the
cosmological expansion, and $\nu$ is from the Jeans diffusion. The
stochastic term $\phi$ is actually the gravitational potential of
the dark matter mass density perturbation, i.e., $\nabla^2\phi=
4\pi G a^2\bar{\rho}_{\rm dm}\delta_{\rm dm}$, and $\delta_{\rm
dm}=(\rho_{\rm dm}-\bar{\rho}_{\rm dm})/\bar{\rho}_{\rm dm}$. The
stochastic property of $\delta_{\rm dm}$ is given by the
correlator $\langle \delta_{\rm dm}({\bf k})\delta_{\rm dm}({\bf
k'})\rangle =P({\bf k}) \delta_{\bf k, k'}$ and the power spectrum
$P({\bf k})\propto k^{-\alpha}$. For the flat cold dark matter
model with a cosmological constant ($\Lambda$CDM), the index is
initially $\alpha=2.3$ in the scale range considered. A basic
feature of the fluid described by Eq.(1) is that the Burgers'
turbulence is developed when the Reynolds number is large
\cite{Po}. For instance, the probability distribution functions of
the velocity differences across a distance $r$, $\delta
v_{r}\equiv \{[{\bf v(x+r)- v(x)}]\cdot {\bf r}/r\}$, are scaling,
when $r$ is larger than the Jeans length \cite{Kim}. These results
indicate that some dynamical features of the cosmic baryon fluid
at low redshifts in the nonlinear regime are turbulencelike.

Therefore, an important question is how well and why the dynamical
state of cosmic baryon fluid at low redshifts can be described as
turbulence. Although the cosmic baryon fluid has been a central
topic in cosmology for a long time \cite{Cen}, no turbulencelike
behavior has been studied. In this Letter, we investigate this
problem in the context of the universal scaling law of fully
developed turbulence. Using numerical samples, we first show that
the universal scaling does exist in the cosmic baryon fluid, and
then analyze the underlying physics of this scaling.

{\it Intermittent exponent.}---The universal property of fully
developed turbulence is measured by the structure functions
$S_p(r)$ and intermittent exponent $\zeta_p$, defined as
\begin{equation}
S_p(r)\equiv \langle \delta v_r^p\rangle \sim r^{\zeta_p}.
\end{equation}
Based on dimensional argument of hierarchical evolution,
Kolmogorov in 1941 predicted that for fully developed turbulence
on scales of inertial range, the intermittent exponent is
$\zeta_p=p/3$ \cite{Ko}. Experimental and numerical results do
not, however, support the $p/3$ law. It should be attributed to
intermittency; i.e., turbulence field in the inertial range is
characterized by stronger non-Gaussianity on smaller scales
\cite{Frisch}. A remarkable development was made by She and
Leveque \cite{SL} (SL hereafter). They proposed that the
non-Gaussian behavior of fully developed turbulence is determined
by the hierarchical structure originated from the Navier-Stokes
equation, and the $p/3$ law should be replaced by
\begin{equation}
\zeta_p/\zeta_3=[1-C(1-\beta^3)]p/3+ C(1-\beta^{p}),
\end{equation}
where $C$ is the Hausdorff codimension of the most dissipative
structures, and parameter $\beta$ is given by the hierarchical
evolution (see below). The SL formula Eq.(3) is in excellent
agreement with various experiments of turbulence \cite{She},
including also turbulence in compressible fluid \cite{MHD}. The SL
scaling law of structure function is considered to be universal
for characterizing the fully developed turbulence.

To investigate the universal scaling of cosmic baryon fluid, we
use the cosmological hydrodynamic simulation samples produced by
the code WIGEON (Weno for Intergalactic medium and Galaxy
Evolution and formatiON) \cite{FLL}. This is a hybrid
hydrodynamic/$N$-body simulation, consisting of the WENO algorithm
\cite{Shu} for baryonic fluid, and $N$-body simulation for
particles of dark matter. The baryon fluid obeys the Navier-Stokes
equation, and is gravitationally coupled with collisionless dark
matter. We use the $\Lambda$CDM cosmological model with parameters
given by the recent observations of the cosmic microwave
background radiation \cite{WMAP}. The linear power spectrum of
mass density perturbations is taken from the fitting formulas of
Eisenstein \& Hu \cite{EH}. The atomic processes in the plasma of
hydrogen and helium of primordial composition, including
ionization, radiative cooling and heating, are modeled in the same
way as in \cite{Cen}.

The simulations were performed in a periodic, cubic box of size 64
$h^{-1}$Mpc with a 512$^3$ grid and an equal number of dark matter
particles. The simulations start at a redshift $z=99$. A uniform
UV-background of ionizing photons is switched on at $z =6 $ to
heat the gas and reionize the universe. The temperature of the
baryonic gas generally lies in the range $10^4-10^6$ K, and the
speed of sound $v_s$ in the baryonic gas is only a few km s$^{-1}$
to a few tens km s$^{-1}$. The Jeans length $\lambda_J$ yields a
term like the viscosity $\mu\simeq v_s\lambda_J$. On the other
hand, the bulk velocity of the baryonic gas is of the order of
hundreds of km s$^{-1}$ \cite{ZF}. Therefore, the Reynolds number
would be larger than $\sim$100 if the scales under consideration
are larger than the Jeans length $\lambda_J$, which is in the
range $\sim 0.1 - 0.3$ h$^{-1}$ Mpc for redshifts $z<4$ \cite{Bi}.

These samples are successful to model the power spectrum and
non-Gaussian features of the observed transmitted flux of
Ly$\alpha$ \cite{FPF}. Hence it would be suitable to study whether
the universal scaling law Eq.(3) is available for cosmic baryon
fluid. For this purpose, we randomly sampled 10,000
one-dimensional sub-samples, with each one containing 512 data
points. At each point of a line, the peculiar velocity and mass
density of the baryonic gas is recorded.

We calculated the moments of velocity difference $\delta v_r$ for
$r$ from 1 to 16 h$^{-1}$ Mpc, of which the lower limit is larger
than the Jeans length, and the upper limit is due to the size of
the simulation box. The intermittent exponent $\zeta_p$ for the
sample at redshift $z=0$ is shown in Fig.1. The error bars are the
variance of the samples. The intermittency of cosmic baryon fluid
is excellently fitted by the scaling Eq.(3) on all orders
considered, if the parameters of SL formula [Eq.(3)] are taken to
be $C=1$ and $\beta^3=1/3$.

Our sample is very different from experimental and numerical
samples \cite{Be} used to test the SL formula. The success shown
in Figure 1 further supports that the hypotheses used to derive
the SL formula Eq.(3) are universal so that it would also be
available for cosmic baryon fluid. In this sense, the non-Gaussian
behavior of the baryon fluid in an expanding universe at redshift
$z=0$ can essentially be described as a fluid satisfying the
universal scaling for fully developed turbulence.

\begin{figure}
\rotatebox{0}{\resizebox{8.5cm}{5.253cm}
{\includegraphics{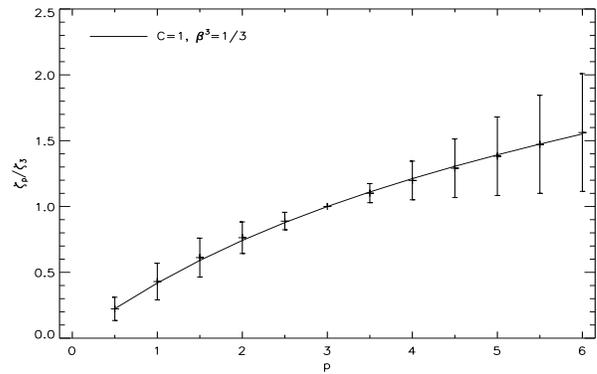}}} 
\caption{Intermittent exponent $\zeta_p$ of cosmic baryon fluid at
         redshift $z=0$. The means and error bars are from the average
         over samples on scale range from $r=1$ to 16 h$^{-1}$ Mpc.
         The solid line is the She-Leveque formula Eq.(2) with $C=1$
         and $\beta^3=1/3$.}
\vspace{-0.4cm}
\end{figure}

\smallskip

{\it Singular dissipative structures.}---Simply speaking, the
underlying physical picture of the SL formula is as follows. In
the inertial range, the fluid evolution is governed by
scale-covariant interactions from the Navier-Stokes equations. The
kinetic energy of fluid is dissipated in singular structures, and
the energy dissipation on different scales satisfies hierarchical
relation.

Although the clustering of baryon fluid is governed by the gravity
of the background mass field, the expansion of the Universe
eliminates the gravity of the uniformly distributed matter. The
peculiar motion of baryon fluid feels only the gravity given by
the random fluctuations of the distributed dark matter. In a
nonlinear regime, baryon fluid decouples from dark matter. It is
similar to the decoupling of a passive substance from the
underlying field during nonlinear evolution \cite{Shra}. The
decoupling leads to the velocity of baryon fluid generally being
less or not larger than the velocity of dark matter \cite{Kim}.
Therefore, baryon fluid is actually driven by a random force of
the dark matter gravity. The potential of the random gravity is of
power law, i.e., scale free. In the nonlinear regime, the power of
fluctuations transfers from larger to smaller scales. Therefore,
in the scale range of $1$ to about 10 h$^{-1}$ Mpc the evolution
of cosmic fluid is similar to turbulence in inertial range.

\begin{figure}
\rotatebox{0}
{\resizebox{4.25cm}{4.25cm}{\includegraphics{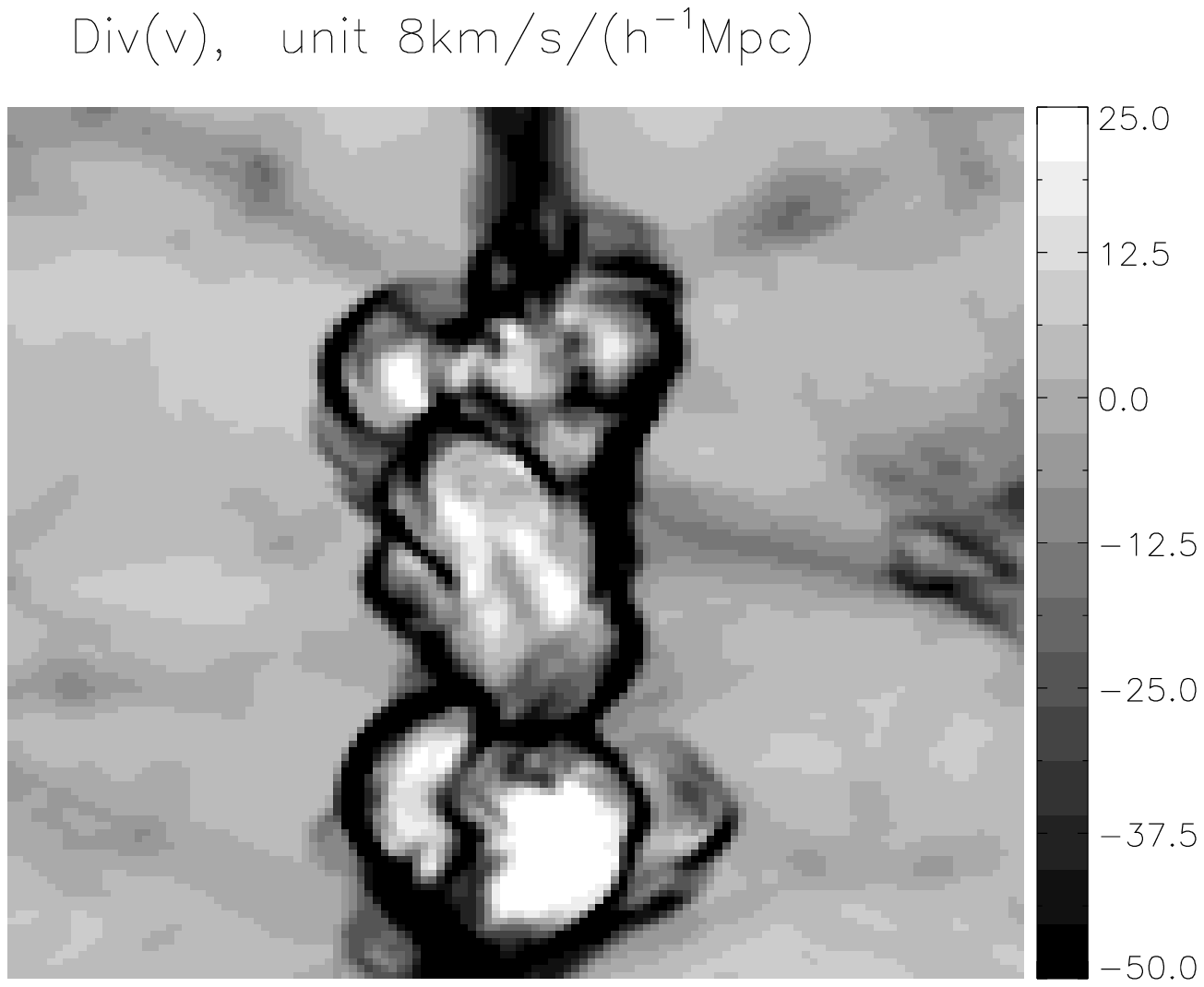}}}
{\resizebox{4.25cm}{4.25cm}{\includegraphics{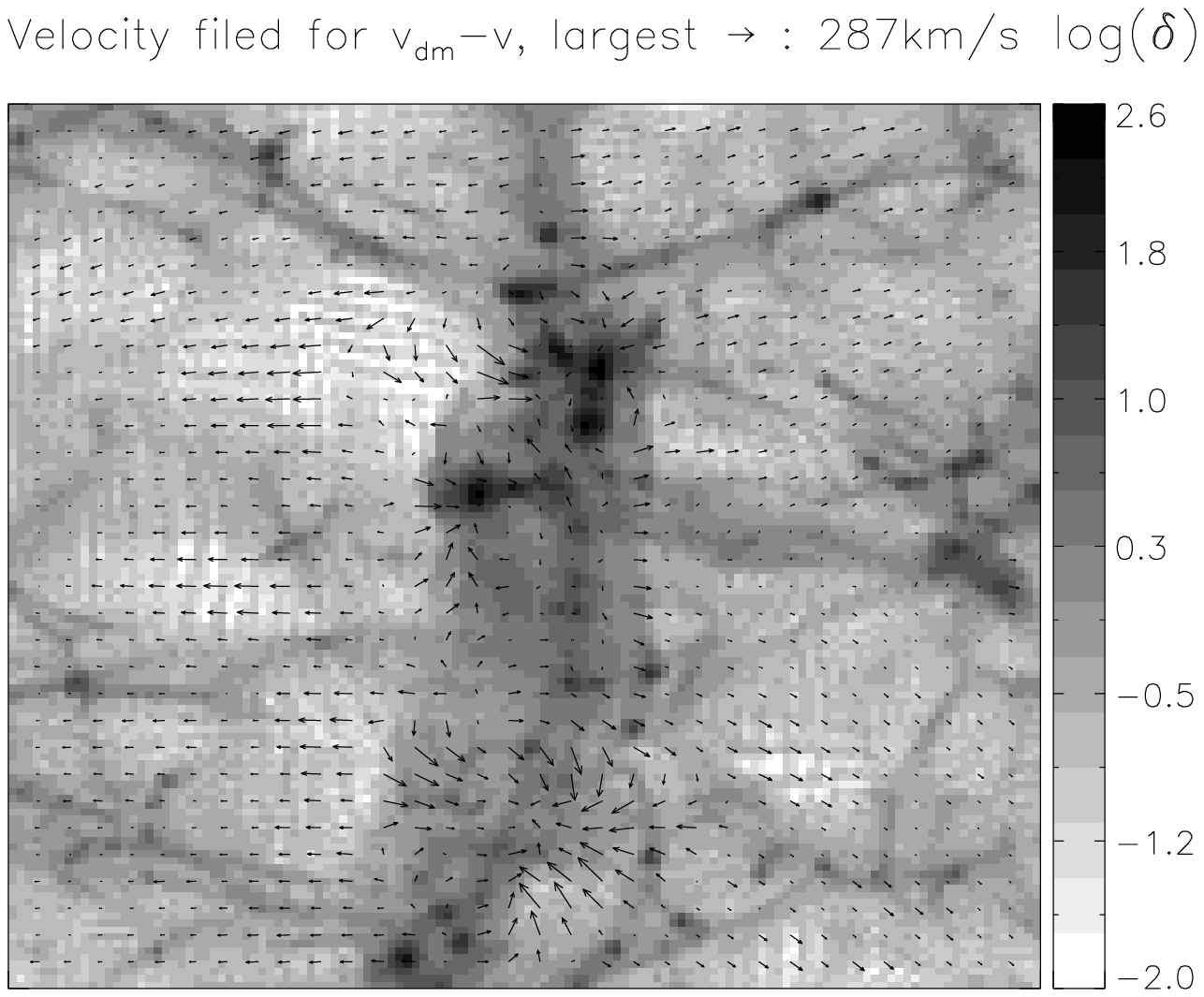}}}
\caption{Left: $\nabla\cdot {\bf v}$ contour of baryonic fluid
         for a slice with area $16\times 16$ h$^{-2}$ Mpc$^2$ and
         thickness 2 h$^{-1}$ Mpc at $z=0$. Right: the field of
         ${\bf v}_{\rm dm}-{\bf v}$ for the same sample as the
         left one. The background is the mass density contrast
         of baryon fluid $\log \delta$.}
\end{figure}

\smallskip

Since $C=1$ gives a best fitting (Fig.1), the singular dissipative
structures should be two-dimensional sheets. It is well known that
in structure formation sheets are dominant in terms of volume
weight statistics. For incompressible fluid, the singular
structures are traced by the amplitude of vorticity $|\nabla
\times {\bf v}|$ \cite{She3}. Since the Burgers turbulence is
caused by shock wave \cite{Po}, the singular dissipative
structures are traced by the divergence of baryon fluid
$\nabla\cdot {\bf v}$. Figure 2 presents a typical $\nabla\cdot
{\bf v}$ contour of baryonic gas in a slice at $z=0$. One can
clearly see the filamentary structures in two dimensions, and
therefore, they are most likely to be sheets in three dimensions.
In Fig.2, we also plot the difference of velocity fields between
dark matter and baryon matter. It shows that the decoupling
between the velocity fields of baryon fluid and dark matter
occurred on the entire field.

{\it Hierarchical relation.}---The parameter $\beta$ of Eq.(3)
describes the hierarchical evolution of the cosmic baryon matter.
The hierarchical clustering is well known in the theory of
large-scale structure formation of the Universe. It has been
proposed that the hierarchical clustering is given by the relation
$\langle \delta\rho_l^N\rangle\propto \langle \delta
\rho_l^2\rangle^{N-1}$, where $\delta\rho_l$ is the density
fluctuation field of the cosmic matter on scale $l$ \cite{White}.
In hydrodynamics, this relation is the so-called pure hierarchical
relation \cite{Du}. It has been shown, however, that the observed
intermittency is not subject to the pure hierarchical relation if
the proportional coefficient between $\langle
\delta\rho_l^N\rangle$ and $\langle \delta \rho_l^2\rangle^{N-1}$
is scale independent \cite{FPF2}.

\begin{figure}
\vspace{0.0cm} \rotatebox{0} {\resizebox{8.5cm}{5.253cm}
{\includegraphics{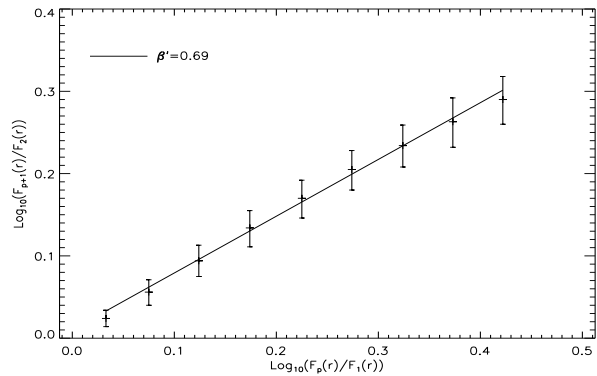}}} 
\caption{The log-log relation between $F_{p+1}(r)/F_2(r)$ and
         $F_{p}(r)/F_1(r)$ for the cosmic baryon fluid sample.
         The slope is $\beta' = 0.69$.}
\vspace{-0.5cm}
\end{figure}

\smallskip

In the universal scaling scheme, the hierarchy is not described by
mass density fluctuation $\delta\rho$, but by a hierarchical
relation of structure functions as \cite{She}
\begin{equation}
F_{p+1}(r)=A_pF_p(r)^{\beta'}F_{\infty}(r)^{1-\beta'},
\end{equation}
where $F_p(r)\equiv S_{p+1}(r)/S_p(r)$, and number $A_p$ are scale
$r$ independent. $F_{p+1}(r)$ describes the intensity of
fluctuations. The larger the $p$ of $F_p(r)$, the higher the
intensity of fluctuations. The most intermittent structures are
described by $F_{\infty}(r)$. Equation (4) describes the
hierarchical relation between fluctuations with different
intensities. Equation (4) is invariant with respect to a
translation in $p$. The parameter $\beta'$, which should be in the
range $0< \beta'<1$, is to measure the degree of intermittency of
turbulent flow. The smaller the $\beta'$, the stronger the
intermittency. For $\beta'=1$, the field is not intermittent.

From Eq.(4), we have
\begin{equation}
\frac{F_{p+1}(r)}{F_2(r)}=\frac{A_p}{A_1} \left (\frac{F_{p}(r)}{
F_1(r)}\right)^{\beta'}.
\end{equation}
If $A_p$ is $p$ independent, we should have
\begin{equation}
\ln F_{p+1}(r)/F_2(r)=\beta'\ln F_{p}(r)/ F_1(r) + {\rm const}.
\end{equation}
Substituting Eq.(3) into Eq.(6), we have $\beta'=\beta$. Equation
(6) does not contain term $F_{\infty}(r)$, and therefore, it can
be directly used to test the assumption of hierarchy. Figure 3
presents the relation of $\log [F_{p+1}(r)/F_2(r)]$ vs $\log
[F_{p}(r)/F_1(r)]$, which is a perfectly straight line. The slope
is $\beta'=0.69\pm 0.02$, which is also perfectly in agreement
with the parameter $\beta=(1/3)^{1/3}=0.69$. Therefore, the
assumption of a constant $A_p$ ($p$ independence) is tenable. The
hierarchical relation Eq.(4) with constant $A_p$ and $\beta=0.69$
provides a good description of the hierarchical structures of
cosmic baryon fluid on the scales of the inertial range.

{\it Discussion and conclusion.}---We showed that all the moments
of the fluctuations of velocity field of cosmic baryon fluid at
$z=0$ on scales larger than the Jeans length have a power-law
dependence on scales, and the intermittent exponents obey the
universal law, which depends only on (1) the dimension of the most
singular dissipative structures and (2) the hierarchical relation
between structures with various intensities of fluctuations. This
result strongly indicates that the highly nonlinear evolution
would lead to the cosmic baryon fluid reaching a statistically
quasiequilibrium state satisfying the universal scaling as that of
fully developed turbulence. It should be emphasized that the
turbulence described by Eq.(1) is stirred at all scales, not only
at the largest scale as in the conventional model of turbulence.
The SL scaling is reasonable not only for the conventional energy
cascade but also the case of Eq.(1).

In view of this picture, we can say that in the highly nonlinear
regime, the statistical properties, especially the intermittent
behavior, of the velocity fluctuations are actually independent of
the details of the dissipative processes. The state depends only
on the dimension of dissipative structures and the hierarchical
relation index. We believe that some statistical features, which
have already been recognized in large-scale structures, would be
directly in consequence of the universal properties of cosmic
baryon fluid.

The velocity and mass density field on scales larger than the
Jeans length are the basic environment of the formation of the
luminous objects. Therefore, the universal properties of the
nonlinear regime provide common frame of studying the dynamical
and thermodynamical evolution of structure formation. For
instance, the redshift dependence of the scaling behavior would be
useful for understanding the evolution of cosmic clustering. The
details will be reported in the near future.

This work is supported in part by the U.S. NSF under Grants No.
AST-0506734 and No. AST-0507340. L.L.F. and P.H. acknowledge
support from the National Science Foundation of China (Grants No.
10573036 and No. 10545002).

\end{document}